\documentclass{birkjour}
\theoremstyle{definition}

\theoremstyle{remark}

\numberwithin{equation}{section}

\usepackage{amsmath}
\usepackage{amssymb}

\begin{document}

%
%
%
%
%
%
%
%
%

\title[]
{Co-Toeplitz Quantization: A Simple Case}

\author[]{Stephen Bruce Sontz}

\address{%
	Centro de Investigaci\'on en Matem\'aticas, A.C. 
	(CIMAT)\\
	Jalisco s/n\\
	Valenciana\\
	Guanajuato 36023\\
	Mexico}

\email{sontz@cimat.mx}




\begin{abstract}
	\noindent 
	The author has introduced in a recent paper
	a new class of operators, called co-Toeplitz 
	operators, with symbols in a co-algebra.
	This is the categorical 
	dual to Toeplitz 
	operators which have symbols in an algebra. 
	The mapping from a symbol to its co-Toeplitz 
	operator gives a quantization scheme, called
	co-Toeplitz quantization. 
	A new, quite simple particular case of co-Toeplitz 
	quantization is introduced in this note. 
	Examples  
	are given in order to show 
	some of its properties. 
\end{abstract}

\maketitle
\tableofcontents
%

\section{Introduction}

In \cite{coT} I have defined co-Toeplitz 
operators in a dual way in terms of category 
theory to Toeplitz operators. 
The structures needed for this definition 
are a co-algebra $ \mathcal{C} $
(see \cite{haze}) together with 
a sesqui-linear form $ \langle \cdot , \cdot \rangle$ 
defined on it. 
We let 
$ \Delta : \mathcal{C}\to\mathcal{C}\otimes\mathcal{C}$ 
denote the co-multiplication 
of $ \mathcal{C} $. 
Also, we suppose there is another 
co-algebra $ \mathcal{P} $ which injects into
$ \mathcal{C} $ by a map 
$ j : \mathcal{P} \to \mathcal{C} $ and that there
is a projection $  Q : \mathcal{C} \to \mathcal{P},  $
that is, $ Q j = id_{\mathcal{P}} $, the 
identity on $ \mathcal{P} $. 
Then we define the {\em co-Toeplitz operator} 
$ C_{g} : \mathcal{P} \to \mathcal{P} $ to be the 
linear operator defined by the composition
\begin{equation}
\label{define-CoToeplitz}
\mathcal{P} \stackrel{j}{\longrightarrow} 
\mathcal{C} \stackrel{\Delta}{\longrightarrow}
\mathcal{C} \otimes \mathcal{C} 
\stackrel{Q \otimes id}{\longrightarrow} 
\mathcal{P} \otimes \mathcal{C} 
\stackrel{\pi_{g}}{\longrightarrow}
\mathcal{P}. 
\end{equation}
The linear map 
$ \pi_{g} : \mathcal{P} \otimes \mathcal{C} \to 
\mathcal{P} $, 
where $ g \in \mathcal{C} $
is called the {\em symbol} of 
the co-Toeplitz operator $ C_{g} $,
is defined in \cite{coT} by
$$
   \pi_{g} ( \phi \otimes f):= \langle g, f \rangle \phi 
$$
for $ \phi \in \mathcal{P}  $ and $ f \in \mathcal{C} $. 
(The map $ \pi_{g} $ is dual to $ \alpha_{g} $ 
to be defined below.)
The {\em anti-linear} map 
$ g \mapsto C_{g} $ is called the 
{\em co-Toeplitz quantization} of $ \mathcal{C} $. 
This in general 
does not involve measure theory. 
For more details see \cite{coT}. 

The definition \eqref{define-CoToeplitz}
 is the dual diagram to that for a 
{\em Toeplitz operator} which is defined 
for a symbol $ g \in \mathcal{A} $, an algebra,  
as the composition (right to left) 
\begin{equation}
\label{define-Toeplitz}
\mathcal{P} \stackrel{P}{\longleftarrow} 
\mathcal{A} \stackrel{\mu}{\longleftarrow}
\mathcal{A} \otimes \mathcal{A} 
\stackrel{\iota \otimes id}{\longleftarrow} 
\mathcal{P} \otimes \mathcal{A} 
\stackrel{\alpha_{g}}{\longleftarrow}
\mathcal{P}. 
\end{equation}
Here $ \mathcal{P} $ is a sub-algebra of 
$ \mathcal{A} $ with 
$ \iota : \mathcal{P} \to \mathcal{A}$ being the
inclusion map and 
$ P : \mathcal{A} \to \mathcal{P}$ being a projection.
Also, $ \mu $ is the multiplication map of 
$ \mathcal{A} $ and 
$ \alpha_{g} (\phi):= \phi \otimes g $ 
for $ \phi \in \mathcal{P}  $. 

The definition \eqref{define-CoToeplitz}
has a particular case:  
$ \mathcal{P} = \mathcal{C} $ and
$ j = Q = id_{\mathcal{C}} $. 
Then 
$$ 
C_{g} = \pi_{g} \, \Delta : \mathcal{C} \to \mathcal{C}. 
$$ 
In this particular simple case, 
which is the new idea in this
note, the only structures needed
are a co-algebra and a sesqui-linear form on it. 
All the examples in this note fall within this case. 
This simple case does not 
have an interesting analogue for Toeplitz operators, 
since diagram \eqref{define-Toeplitz} reduces to 
the right regular representation of 
$ g $ acting on $ \mathcal{A} $
if we put $ \mathcal{P} = \mathcal{A} $. 

If the sesqui-linear form is positive definite, then
$ C_{g} $ acts in a pre-Hilbert space and so may be
considered as a densely defined operator 
acting in the Hilbert space 
completion of $ \mathcal{C} $.  
Thus we can construct models for quantum physics, 
including 
creation and annihilation co-Toeplitz operators. 
(See \cite{coT}.)

All objects in this paper  
are vector spaces over the
complex numbers, and all arrows 
are linear maps, except as noted.

\section{Manin Quantum Plane}

We define $ \mathcal{C} $
to be the algebra generated by two 
elements $ a,c $ with the relation
$ ac = q ca $ for some non-zero $ q 
\in \mathbb{C}$, the complex numbers. 
This is called 
 the {\em Manin quantum plane}. 
The notation follows that used in \cite{coT}. 
We define the co-multiplication $ \Delta $ to be the 
algebra morphism determined by
$$ 
\Delta (a) = a \otimes a \quad  \mathrm{and} \quad  
 \Delta (c) = c \otimes a. 
$$
This is well defined on $ \mathcal{C} $, since 
$ \Delta ( ac - q ca ) =0 $ as the reader can verify. 
We note that 
$ \mathcal{C} $ does not have a co-unit, though 
this has no great importance for our purposes. 
Clearly, 
$\mathcal{B} :=  
\{ a^{i} c^{j}  ~|~ i,j \in \mathbb{N}\} $, 
is a Hamel basis of $ \mathcal{C} $, 
where $ \mathbb{N} $ denotes the non-negative integers. 
Since 
$ C_{g} $ is anti-linear 
in the symbol $ g \in \mathcal{C} $, 
it suffices to calculate $ C_{g} $ for the basis 
elements $ a^{i} c^{j} $. 
And since $ C_{g} $ is linear, it suffices to 
evaluate it on these basis elements. 
We proceed to do this. 
First, we see that 
$$
\Delta (a^{k} c^{l}) = \big( \Delta (a) \big)^{k} 
\big( \Delta (c) \big)^{l} 
= (a \otimes a)^{k} (c \otimes a )^{l} 
= a^{k} c^{l} \otimes a^{k+l}. 
$$
Then
$$
C_{a^{i} c^{j}} (a^{k} c^{l}) = 
\pi_{a^{i} c^{j}} \, \Delta (a^{k} c^{l}) 
= \pi_{a^{i} c^{j}} 
\big( a^{k} c^{l} \otimes a^{k+l} \big)
= \langle a^{i} c^{j} , a^{k+l} \rangle \, a^{k} c^{l}. 
$$
So $ C_{a^{i} c^{j}} $ is diagonalized by the basis
$ \mathcal{B} $ with its eigenvalues determined by 
the sesqui-linear form, and therefore it is neither 
a creation nor an annihilation operator. 
Rather   $ C_{a^{i} c^{j}} $ is what is known 
as a {\em preservation operator}.

At this point in the calculation it 
becomes clear that the definition of the 
sesqui-linear form enters in a fundamental way, namely, 
different choices for it give different 
co-Toeplitz quantizations. 
One simple choice is to choose it so that 
the basis $ \mathcal{B} $ is orthogonal. 
So we put
$$
\langle a^{i} c^{j} , a^{k} c^{l} \rangle :=
\delta_{i,k} \, \delta_{j,l} \, w(i,j).
$$
Here $ \delta_{r,s} $ denotes the Kronecker delta. 
We also take
$ w(i,j) > 0 $ for all $ i,j \in \mathbb{N} $ so that
this is a positive definite inner product and
$ \mathcal{C} $ is a pre-Hilbert space. 
With this inner product we see that 
$$
C_{a^{i} c^{j}} (a^{k} c^{l}) = 
\langle a^{i} c^{j} , a^{k+l} \rangle \, a^{k} c^{l} 
= \delta_{i,k+l} \, \delta_{j,0} \, w(i,j) \, a^{k} c^{l}. 
$$
So $ C_{a^{i} c^{j}} = 0 $ if $ j > 0 $. 
Continuing with the case $ j = 0 $ we see that 
$$
C_{a^{i} } (a^{k} c^{l}) 
= \delta_{i,k+l} \, w(i,0) \, a^{k} c^{l}. 
$$
So $ C_{a^{i} } $ is diagonalized by the basis 
$ \mathcal{B} $ with $ 0 $ and $ w(i,0) $ 
being its eigenvalues. 
We define the degree of the monomial 
$ a^{k} c^{l} $ by $ \deg \, a^{k} c^{l} := k+l $.
Then $ C_{a^{i}} $ is zero on monomials 
with degree $ \ne i $ and is a non-zero 
multiple of the
identity on the finite dimensional 
vector space spanned by 
the monomials of degree~$ i $. 

Another choice for the sesqui-linear form is
$$
\langle a^{i} c^{j} , a^{k} c^{l} \rangle := 
\delta_{i-j,k-l} \, \mu (i, j, k, l), 
$$
where $ \mu : \mathbb{N}^{4} \to (0, \infty)$ is 
a positive weight function. 
With this choice of sesqui-linear form 
we have 
$$
C_{a^{i} c^{j}} (a^{k} c^{l}) = 
\langle a^{i} c^{j} , a^{k+l} \rangle \, a^{k} c^{l} 
= \delta_{i-j,k+l} \, \mu (i,j,k+l,0) \, a^{k} c^{l}.  
$$
Thus the eigenvalues of $ C_{a^{i} c^{j}} $ are $ 0 $
and $\mu (i,j,i-j,0)$. 
Moreover, $ C_{a^{i} c^{j}} $ is zero except on 
the set of 
monomials of degree $ i-j $. 
So $ i < j $ implies that $ C_{a^{i} c^{j}} = 0 $.

\section{Divided Power Co-algebra}

This is based on Example 2.4.8 in \cite{haze}. 
We let $ \mathcal{C} $ be the vector space 
with basis $\{ x_{i} ~|~ i \in \mathbb{N} \}$. 
The co-multiplication $ \Delta $ is the linear map
determined by
\begin{equation}
\label{divided-co-mult}
   \Delta (x_{n}):= \sum_{i+j = n} x_{i} \otimes x_{j}. 
\end{equation}
The degree of each basis element is defined by 
$ \deg x_{n}:= n $. 
We also define a sesqui-linear form by
\begin{equation}
\label{divided-inner-product}
\langle x_{i} , x_{j} \rangle := w(i) \, \delta_{i,j}, 
\end{equation}
where $ w : \mathbb{N} \to (0,\infty) $ is a 
strictly positive weight function. 
So this is an inner product making $ \mathcal{C} $ 
into a pre-Hilbert space. 
Again, it suffices to 
compute the co-Toeplitz operators $ T_{g} $
for $ g $ in the basis. 
So we compute as follows: 
\begin{align*}
C_{x_{k}} (x_{n}) &= \pi_{x_{k}} \Delta (x_{n}) 
= \pi_{x_{k}} 
\big( \!\! \sum_{i+j = n} \!\! x_{i} \otimes x_{j} \big) 
\\
&= \sum_{i+j = n} \langle  x_{k} , x_{j} \rangle \, x_{i} 
= \sum_{i+j = n} \delta_{k,j} w(k) \, x_{i}.  
\end{align*}
Now if $ k > n $ we have $ \delta_{k,j} = 0 $ for
all the terms in the last sum, since $ j \le n $. 
So $ C_{x_{k}} (x_{n}) = 0 $ if $ k > n $. 
For the opposite case $ 0 \le k \le n $ we have 
$$
C_{x_{k}} (x_{n}) = 
\sum_{i+j = n} \delta_{k,j} w(k) \, x_{i} = 
w (k) \, x_{n-k}. 
$$
If we define $ x_{i} := 0 $ and $ w(i):= 0 $
for all integers $ i < 0 $, then 
we can write this result as one formula for 
all $ k,n \in  \mathbb{N}$: 
$$
C_{x_{k}} (x_{n}) = w (k) \, x_{n-k}. 
$$
So for $ k > 0 $ we have that $ C_{x_{k}} $
decreases degree by $ k $ and so is an annihilation 
operator. 
On the other hand 
$ C_{x_{0}} $ is a preservation operator.

\section{Negative Degrees}

Here we give a modification of the previous example
that includes negative degrees. 
We let $ M \ge 1 $ be an integer and define
$\mathcal{C}$
to be the complex vector space with basis
$ \{ x_{i} \}$ for integers $ i \in [-M, M] $. 
So $ \dim \mathcal{C} = 2M+1 $. 

We define $ \deg x_{i}:= i $ for  $ i \in [-M, M] $. 
For convenience we also define $ x_{i}:= 0 $ for all 
integers $ i $ with $ |i| > M $. 
We use the same formulas as in the previous 
example, but with new interpretations. 
So, 
the co-multiplication $ \Delta $ is defined by
\eqref{divided-co-mult}, but now 
for integers $ |n| \le M $.  
With our definitions only finitely many terms in 
the (now) infinite 
sum \eqref{divided-co-mult} are non-zero. 
We also define a sesqui-linear form by 
\eqref{divided-inner-product} but now for integers 
$ i,j \in [-M , M] $. 
Again, 
for convenience we put $ w(i):= 0 $ for $ |i| > M $. 
The same calculation as in the previous example gives
$$
C_{x_{k}} (x_{n}) = w (k) \, x_{n-k} 
$$
but now for all integers $ k,n \in [-M , M] $. 
There are three cases: 
\begin{enumerate}
	\item   
		$ C_{x_{k}} $ increases degree by $ |k| $ 
	if $ k < 0 $ and is a creation operator. 
 	\item   
 	$ C_{x_{k}} $ decreases degree by $ k $ 
 	if $ k > 0 $ and is an annihilation operator. 
 	\item 
 	$ C_{x_{k}} $ preserves degree if $ k =0 $ 
 	and is a preservation operator. 
\end{enumerate}

So we get the three types of operators relevant 
to physics by using the basis elements with positive, 
negative and zero degrees. 
We also can define a {\em $ * $-operation} 
($ \equiv $ {\em conjugation})
on $ \mathcal{C} $ 
by putting $ x_{i}^{*}:= x_{-i} $. 
This definition is motivated by the theory 
of complex variables. 
Using this as motivation,  
for each $ i > 0 $ we then define the elements $ x_{i} $ 
to be {\em holomorphic} and the elements 
$ x_{i}^{*} $ to be {\em anti-holomorphic}. 
(The element $ x_{0} $ could be defined as
being both holomorphic and anti-holomorphic,
if one wished. 
But I opt not to do that.)
Then the anti-holomorphic elements are symbols of 
creation operators while the holomorphic 
elements are symbols of annihilation operators.

\section{Matrix Coalgebra}

This example comes from Example 2.4.1 
in \cite{haze}. 
Let $ \mathcal{C} $ be the vector space with 
basis $\{ E_{i,j} ~|~ 1 \le i, j \le n \} $, where
$ n \ge 1 $ is an integer. 
So, $ \dim \, \mathcal{C} = n^{2} $. 
Of course, the motivation is that $ E_{i,j} $ 
is analogous to the $ n \times n $ matrix with
all entries $ 0 $, except for row $ i $ and column $ j $ 
which has the entry $ 1 $.
Let the co-multiplication be determined by
$$
    \Delta (E_{i,j}):= 
    \sum_{k=1}^{n} E_{i,k} \otimes E_{k,j}. 
$$
(As a curious parenthetical remark, 
let us note that this 
vector space has a natural algebra structure
motivated by matrix multiplication. 
However, this does not combine with this 
co-multiplication to give us a bi-algebra.
See \cite{haze}.)
We define an inner product on $ \mathcal{C} $
by making the basis $\{ E_{i,j} \}$ 
orthonormal. 
Then we calculate 
\begin{align*}
&C_{E_{r,s}} (E_{i,j}) = \pi_{E_{r,s}} \, \Delta (E_{i,j})
= \pi_{E_{r,s}} 
\big( \sum_{k=1}^{n} E_{i,k} \otimes E_{k,j} \big) 
\\
&
=  \sum_{k=1}^{n} 
\langle E_{r,s} ,  E_{k,j} \rangle E_{i,k}
=  \sum_{k=1}^{n} 
\delta_{r,k} \delta_{s,j}
E_{i,k}
= \delta_{s,j} E_{i,r}. 
\end{align*}
So, $ C_{E_{r,s}} (E_{i,j})$ is either zero or 
another basis element. 

Another sesqui-linear form is given by
$
\langle E_{i,j} , E_{r,s} \rangle := 
w (i+s) \,  \delta_{i-j,r-s} 
$
with a weight function 
$ w : \mathbb{N} \to (0, \infty) $. 
Then  
\begin{align*}
&C_{E_{r,s}} (E_{i,j}) = \pi_{E_{r,s}} \, \Delta (E_{i,j})
= \pi_{E_{r,s}} 
\big( \sum_{k=1}^{n} E_{i,k} \otimes E_{k,j} \big) 
\\
&
=  \sum_{k=1}^{n} 
\langle E_{r,s} ,  E_{k,j} \rangle E_{i,k}
=  \sum_{k=1}^{n} 
w (r+j) \delta_{r-s,k-j} \, E_{i,k}
= w (r+j) E_{i,j+r-s}, 
\end{align*}
where we put $ E_{i,j} =0 $ if $ j \le 0 $ or $ j > n $. 
If we define $ \deg E_{i,j} := i+j $, then we see 
that $ C_{E_{r,s}} $ changes degree by $ r-s $. 
So, $ C_{E_{r,s}} $ is a creation operator if $ r > s $, 
it is an annihilation operator if $ r< s $ and 
finally it is a preservation operator if $ r =s $. 

Using the adjoint operation of matrices as 
motivation, 
we also define a $ * $-operation by 
$ E_{i,j}^{*} := E_{j,i}$. 
We also say that $ E_{i,j} $ is {\em holomorphic} 
if $ i < j $ 
(`upper triangular')
and is  {\em anti-holomorphic} 
if $ i > j $ 
(`lower triangular'). 
As previously, 
the anti-holomorphic $ E_{i,j} $ are the symbols of 
(degree increasing)
creation operators and, on the other hand, 
the holomorphic $ E_{i,j} $ 
are the symbols of 
(degree decreasing)
annihilation operators. 
Also 
the `diagonal' elements $ E_{i,i} $, which are
self-adjoint (or real) 
with respect to the $ * $-operation, 
are symbols of (degree preserving)
preservation operators.

\section{Concluding Remarks}

The quantization of co-algebras is a new 
field of research with 
co-Toeplitz quantization being the 
first theory that achieves this. 
It is remarkable that any sesqui-linear form 
defined on a co-algebra $ \mathcal{C} $
is sufficient extra structure 
to give us a co-Toeplitz quantization 
of $ \mathcal{C} $. 
It is noteworthy that in some of these examples 
a $ * $-operation can be defined thereby 
giving holomorphic 
and anti-holomorphic elements, which are symbols whose  
co-Toeplitz operators are annihilation and creation 
operators, respectively.

\vskip 0.2cm
\centerline{\bf Acknowldgement} 
\vskip 0.2cm \noindent 
I thank the organizers of the XXXVII Workshop on Geometric
Methods in Physics held in July, 2018 
in Bia\l owie$\dot{\rm z}$a, Poland 
for the
opportunity to present a talk  
based on my paper 
\cite{coT} and for inviting me to write this 
note which furthers that continuing research program.

\end{document}